%% file: final_revised.tex
\documentclass{article}
\usepackage{spconf,amsmath,epsfig}
\usepackage{amsfonts}
\usepackage{amsmath}
\usepackage{epsfig}
\usepackage{mathrsfs}
\usepackage{graphicx}
\usepackage{amssymb}

\newtheorem{theorem}{Theorem}

\newtheorem{definition}{Definition}

\newtheorem{corollary}{Corollary}

\newcommand{\qedsymbol}{\hspace{\fill}\rule{1.5ex}{1.5ex}}
\input{tcilatex}

\def\b0{\mbox{\boldmath $0$}}

\def\by{\mbox{\boldmath $y$}}

\def\buno{\mbox{\boldmath $1$}}

\begin{document}
\ninept
\title{\vspace{-0.9cm}Distributed Decision Through
Self-Synchronizing Sensor Networks in the Presence of Propagation
Delays and Nonreciprocal Channels\vspace{-0.2cm}} \name{Gesualdo
Scutari, Sergio Barbarossa and Loreto Pescosolido\vspace{-0.4cm}}
%EndAName
\address{\small Dpt. INFOCOM, Univ. of Rome \textquotedblleft La
Sapienza \textquotedblright, Via Eudossiana 18, 00184 Rome, Italy\\
\small E-mail: \texttt{$\{$scutari, sergio, loreto$\}$@infocom.uniroma1.it.%
\thanks{%
This work has been partially funded by the WINSOC project (Contract
N0033914), and by ARL/ERO (Contract N62558-05-P-0458).}}}
\maketitle\vspace{-2.1cm}

\begin{abstract}
In this paper we propose and analyze a distributed algorithm for
achieving globally optimal decisions, either estimation or
detection, through a self-synchronization mechanism among linearly
coupled integrators initialized with local measurements. We model
the interaction among the nodes as a directed graph with weights
dependent on the radio interface and we pose special attention to
the effect of the propagation delays occurring in the exchange of
data among sensors, as a function of the network geometry. We derive
 necessary and sufficient conditions for the proposed system to
reach a consensus on globally optimal decision statistics. One of
the major results proved in this work is that a consensus is
achieved for any bounded delay condition if and only if the directed
graph is quasi-strongly connected. We also provide a closed form
expression for the global consensus, showing that the effect of
delays is, in general, to introduce a bias in the final decision.
The closed form expression is also useful to modify the consensus
mechanism in order to get rid of the bias with minimum extra
complexity.
\end{abstract}\vspace{-0.2cm}

\section{Introduction and Motivations}
\vspace{-0.1cm}
Endowing a sensor network with self-organizing capabilities is
undoubtedly a useful goal to increase the resilience of the network
against node failures (or simply switches to sleep mode) and avoid
potentially dangerous congestion conditions around the sink nodes.
Decentralizing decisions decreases also the vulnerability of the
network against damages to the sink or control nodes. Distributed
computation over a network has a long history, starting with the
pioneering work of Tsitsiklis, Bertsekas and Athans
\cite{Tsitsiklis-Bertsekas-Athans} on asynchronous agreement problem
for discrete-time distributed decision-making systems and parallel computing
\cite{Tsitsiklis-Bertsekas-book}. A simple, yet
significant, form of in-network distributed computing is achieving a consensus
about one common observed phenomenon, without the presence of a
fusion center. Linear average consensus techniques have received great attention in the
recent years \cite{Olfati-Saber-2003}$-$\cite{Chai-Wu-2005}. An excellent
tutorial on distributed consensus techniques is given in
\cite{Olfati-Saber-Murray-ProcIEEE07}.
\\\indent Consensus may be also seen as a form of self-synchronization among
coupled dynamical systems.
In \cite{Barbarossa-iwwan05, Barbarossa-Scutari-Journal}, the authors
showed how to use the self-synchronization capabilities of a set of
nonlinearly coupled first-order
dynamical systems to reach the \textit{%
global} maximum likelihood estimate, assuming reciprocal
communication links. In particular, in \cite{Barbarossa-Scutari-Journal} it was shown that reaching a consensus on the
state derivative, rather than on the state itself (as in \cite{Tsitsiklis-Bertsekas-Athans}$-$\cite{Olfati-Saber-Murray-ProcIEEE07}),
allows for better
resilience against coupling noise.\\ %In
%\cite{Schizas-Ribeiro-Giannakis}, a distributed algorithm to compute
%the  best linear unbiased estimator of a deterministic parameter
%vector based on consensus was proposed.
\indent The consensus protocols proposed in
\cite{Olfati-Saber-2003}$-$\cite{Olfati-Saber-Murray-ProcIEEE07}
assume that the interactions among the nodes occur instantaneously,
i.e., without any propagation delay. However, this assumption is not
valid for large scale networks, where the distances among the nodes
are large enough to introduce a nonnegligible communication delay.
%Most works on consensus assume that the interaction among nodes
%occur instantaneously, i.e., there is no propagation delay.
There are only a few recent works that studied the consensus problem
for \emph{time-continuous} dynamical systems
in the presence of propagation delays \cite{Olfati-Saber},
\cite{Strogatz}$-$\cite{Scutari-Barbarossa-Delay-SPAWC}.
The discrete-time case was addressed in \cite{Tsitsiklis-Bertsekas-Athans},
\cite[Ch. 7.3]{Tsitsiklis-Bertsekas-book}, where the authors studied alternative asynchronous linear agreement \emph{discrete-time} algorithms. In
particular, \cite{Olfati-Saber, Strogatz} provided sufficient
conditions for the convergence of a linear consensus protocol in the
case of time-invariant \emph{homogeneous} delays (i.e., equal delay
for all the nodes) and assuming \emph{reciprocal} communication
links. The most appealing feature of the dynamical system in
\cite{Olfati-Saber} is the convergence of the state variables to a
common value, which is \emph{known} in advance (equal to the
weighted average of the initial conditions) and
\emph{delay-independent}. Unfortunately, this desired property is
paid in terms of convergence capabilities, since, in the presence of
homogeneous delays, the system in \cite{Olfati-Saber} is able to
reach a consensus if and only if the delay is smaller than a given,
topology-dependent, value. %Moreover, the assumptions of homogeneous
%delays and reciprocal channels are not appropriate for describing
%the propagation in a common network deploying scenario.\\%, where in
%%general there are \emph{nonhomogeneous} delays and
%%\emph{nonreciprocal} communication channels.
 The protocol of \cite{Olfati-Saber} was generalized in
\cite{Papachristodoulou-CDC06, Scutari-Barbarossa-Delay-SPAWC} and
\cite{Lee-Spong-06} to the case of time-invariant
\emph{nonhomogeneous} delays (but reciprocal channels) and
nonreciprocal channels, respectively. The dynamical systems studied
in \cite{Papachristodoulou-CDC06, Lee-Spong-06} are guaranteed to
reach a consensus for any given set of finite propagation delays
(provided that the network is strongly connected), but this common
value is not related to the initial conditions of the system by a
\emph{known} function. Similar results, under weaker (sufficient) conditions on the (possibly time-varying) network topology, were obtained in \cite[Ch. 7.3]{Tsitsiklis-Bertsekas-book} for the convergence of discrete-time asynchronous agreement algorithms. This means that the final global consensus
achievable by the systems in \cite{Papachristodoulou-CDC06,
Lee-Spong-06} and \cite[Ch. 7.3]{Tsitsiklis-Bertsekas-book} is not predictable a priori, so that the protocols in
the cited works cannot be used to
distributively compute prescribed functions of the sensors'
measurements, like decision tests or global parameter estimates.\\
\indent Ideally, we would like to have a totally decentralized
system that reaches a global consensus, for \emph{any} given set of
\emph{nonhomogeneous} propagation delays (as in
\cite{Papachristodoulou-CDC06}, \cite[Ch. 7.3]{Tsitsiklis-Bertsekas-book}) and \emph{nonreciprocal} channels
(as in \cite{Lee-Spong-06}, \cite[Ch. 7.3]{Tsitsiklis-Bertsekas-book}), whose final value is a \emph{known} and
\emph{delay-independent} function of the sensors' measurements (as
in \cite{Olfati-Saber}). In this paper we fill this lack and propose
a distributed time-continuous dynamical system having all the above desired
features. More specifically, we consider a set of linearly coupled
first-order dynamical systems, in a network with arbitrary time-invariant topology
(not necessarily strongly connected, as opposed to
\cite{Strogatz}$-$\cite{Scutari-Barbarossa-Delay-SPAWC}) and
nonreciprocal communication channels, modeled as a weighted directed
graph with weights dependent on the physical radio channels. The
links among the nodes are affected by time-invariant nonhomogeneous
time offsets, taking into account the propagation delays,
proportional to the relative distances among the nodes, and clock
offsets among the nodes. Our main contributions are the following: 
i) We provide necessary and sufficient conditions ensuring local 
or global convergence of our dynamical system; ii) We
derive the closed form expression for the consensus, as a function
of the attenuation coefficient and propagation delay corresponding
to each link among the sensors; iii) We show how to get a final
estimate that is not biased by the network geometry and coincides
with the globally optimal decision statistics that would have been
computed by a centralized network having a fusion center that has
ideal access to all the nodes. The most appealing feature of the
proposed system is the convergence of the state derivatives to a
common \emph{known} value, %which represents a globally optimal
%decision statistics,
for {\it any} given set of propagation delays
and nonreciprocal communication channels, with the only requirement
that the network be quasi-strongly connected.
\vspace{-0.2cm}

\section{Reaching Consensus Through Self-Synchronization}
\vspace{-0.2cm}
\label{Problem formulation}% It was recently shown in
%\cite{Giridhar-Kumar05} that, in many applications, an efficient way
%to design a sensor network should follow the so called
%\textit{data-centric} viewpoint, according to which the goal of the
%network is to compute a function of the measurements, where what is
%important is the measurement per se, and not necessarily which node
%has taken which measurement. 
In this section, we first show a class of
functions that can be computed with a distributed approach and then
we illustrate the mechanism to carry out the computation without the
need of any fusion center.\vspace{-0.3cm}

\subsection{Consensus Achievable with a Decentralized Approach}
\label{SUb-Sec_consensus_examples} If we denote by $y_i$, $i=1,
\ldots, N$ the (scalar) measurement taken from
node $i$, in a network composed of $N$ nodes, we have shown in \cite%
{Barbarossa-Scutari-Journal} that it is possible
to compute any function of the collected data expressible in the
form\vspace{-0.3cm}
\begin{equation}
f(y_{1},y_{2},\ldots ,y_{N})=h\left[ \frac{\dsum_{i=1}^{N}c_{i}g_{i}(y_{i})}{%
\dsum_{i=1}^{N}c_{i}}\right] ,  \label{f}
\end{equation}%
where $\left\{ c_{i}\right\} $ are positive coefficients and
$\{g_{i}\}$ and $h$ are arbitrary (possibly nonlinear) real
functions on $\mathbb{R}$, and  i.e., $g_{i},h:%
%TCIMACRO{\U{211d} }%
%BeginExpansion
\mathbb{R}
%EndExpansion
\mapsto
%TCIMACRO{\U{211d} }%
%BeginExpansion
\mathbb{R}
%EndExpansion
$, in a totally decentralized way, i.e. without the need of a sink node. In
the vector observation case, the function may be generalized to the vector
form
\begin{equation}
\mathbf{f}(\mathbf{y}_{1}\mathbf{,y}_{2}\mathbf{,\ldots ,y}_{N})=\mathbf{h}%
\left[ \left( \sum_{i=1}^{N}\mathbf{C}_{i}\right) ^{-1}\left( \sum_{i=1}^{N}%
\mathbf{C}_{i}\boldsymbol{g}_{i}\mathbf{(y}_{i}\mathbf{)}\right)
\right] , \label{f_vect}
\end{equation}%
where $\{\boldsymbol{g}_{i}\}$ and $\mathbf{h}$ are arbitrary
(possibly
nonlinear) real functions on $\mathbb{R}^{L}$, i.e., $\boldsymbol{g}_{i},\mathbf{h}:%
%TCIMACRO{\U{211d} }%
%BeginExpansion
\mathbb{R}
%EndExpansion
^{L}\mapsto
%TCIMACRO{\U{211d} }%
%BeginExpansion
\mathbb{R}^L
%EndExpansion
$, and $\left\{ \mathbf{C}_{i}\right\} $ are arbitrary square positive
definite matrices.

Even though the class of functions expressible in the form (\ref{f}) or (\ref%
{f_vect}) is not the most general one, it includes many cases of
practical interest, like, e.g., the computation of hypothesis
testing problem, the linear ML estimation
\cite{Barbarossa-Scutari-Journal, Scutari-Barbarossa-Delay-J}, the
sufficient statistic in detection of Gaussian processes in Gaussian
noise, the maximum, the minimum, the
histograms, the geometric mean of the sensors' measurements, and so
on \cite{Barbarossa-Scutari-Journal,
Scutari-Barbarossa-Delay-J}.\vspace{-0.3cm}

\subsection{How to Achieve the Consensus in a Decentralized Way}

The next, most interesting question is how to achieve the aforementioned
optimal statistics in a totally decentralized network. % In \cite%
%{Barbarossa-iwwan05, Barbarossa-Scutari-Journal} we proposed an
%approach to solve this problem using a nonlinear interaction model
%among the nodes, based on an undirected graph and with no
%propagation delays in the exchange of information among the sensors.
In this paper, we consider a linear interaction model among the
nodes, and we generalize the approach of \cite%
{Barbarossa-Scutari-Journal} to a network where
the propagation delays are taken into account and the network is
described through a weighted \emph{directed} graph (or
\emph{digraph,} for short), which is a more appropriate model to
capture the nonreciprocity of the communication links governing the
interaction among the nodes.

The proposed sensor network is composed of $N$ nodes, each equipped
with four basic components: i) a \textit{transducer} that senses the
physical parameter of interest (e.g., temperature, concentration of
contaminants, radiation, etc.); ii) a \textit{local processing unit}
that processes the measurement taken by the node; iii) a
\textit{dynamical system} whose state evolves according to a
first-order differential equation, initialized with the local
measurements, whose state evolves interactively with the states of
nearby sensors; iv) a \textit{radio interface} that transmits the
state of the dynamical system and receives the state transmitted by
the other nodes, thus ensuring the interaction among nearby nodes.

In the scalar observation case, the dynamical system present in node
$i$ evolves according to the following functional differential
equation\vspace{-0.3cm}
\begin{equation}
\begin{array}{l}
\dot{{x}}_{i}(t)=g_i(y_i)+\dfrac{K}{c_{i}}\dsum\limits_{j\in \mathcal{N}%
_{i}}^{N}a_{ij}\,\left( x_{j}(t-\tau _{ij})-x_{i}(t)\right) ,\quad \,\,t>0,
\\
x_{i}(t)={\phi }_{i}(t),\quad t\in \lbrack -\tau ,\ 0],\quad
i=1,\ldots ,N,
\end{array}%
%\begin{array}{l}
%\,\,i=1,\ldots ,N,%
%\end{array}
\label{linear delayed system}
\end{equation}%
where $x_{i}(t)$ is the state function associated to the $i$-th
sensor; $g_{i}(y_{i})$ is a function of the local observation
$y_{i}$; $K$ is a positive coefficient measuring the global coupling
strength; $c_{i}$ is a positive coefficient that may be adjusted to
achieve the desired consensus; $\tau _{ij}=T_{ij}+d_{ij}/c$ is a
delay incorporating the propagation delay due to traveling the
internode distance $d_{ij}$, at the speed of light $c$, plus a
possible time offset $T_{ij}$. The sensors are assumed to be fixed
so that all the delays are constant. We also assume, realistically,
that the maximum delay is bounded, with maximum value $\tau
=\max_{ij}\tau _{ij}.$ The coefficient $a_{ij}$ measures the
amplitude of the signal received
from node $i$ and transmitted from node $j$. We assume that the radio interface is such that $a_{ij}=\sqrt{%
P_{j}|h_{ij}|^{2}/d_{ij}^{\eta }}$, where $P_{j}$ is the power of
the signal transmitted from node $j$; $h_{ij}$ is a fading
coefficient describing the channel between nodes $i$ and $j$; $\eta
$ is the path loss exponent. This requires some form of channel compensation at the receiver side, if the coefficients $h_{ij}$'s are complex. Furthermore, we assume, realistically,
that node $i$ \textquotedblleft hears\textquotedblright\ node $j$
only if the power received from $j$ exceeds a given threshold. In
such a case, $a_{ij}\neq 0$, otherwise $a_{ij}=0$. The set of nodes
that sensor $i$ hears is denoted by $\mathcal{N}_{i}=\{j=1,\ldots
,N:a_{ij}\neq 0\}.$ \ Observe that, in general, $a_{ij}\neq a_{ji}$,
i.e. the channels are non-reciprocal.

Because of the delays, the state evolution (\ref{linear delayed
system}) for, let us say, $t>0$, is uniquely defined provided that
the initial state variables $x_i(t)$ are specified in the interval
from $-\tau $ to $0$. The initial conditions of (\ref{linear delayed
system}) are assumed to be taken in the set of continuous bounded
functions ${\phi }_{i}(t) $
mapping the interval $[-\tau ,\ 0]$ to $%
%TCIMACRO{\U{211d} }%
%BeginExpansion
\mathbb{R}
%EndExpansion
$.

Because of the space limitation, in this paper we focus only on the
case of scalar observations from the sensors. However, similar
results can be generalized to the vector case
\cite{Scutari-Barbarossa-Delay-J}.
\vspace{-0.2cm}
\subsection{Self-Synchronization}

Differently from most papers dealing with average consensus problems
\cite{Tsitsiklis-Bertsekas-Athans}$-$\cite{Olfati-Saber-Murray-ProcIEEE07}
, where the global consensus was intended to be the situation where
all dynamical systems reach the same \emph{state} value, we adopt
here the alternative definition already introduced in our previous
work \cite{Barbarossa-Scutari-Journal}. We define the network
synchronization (consensus) with respect to the state
\textit{derivative}, rather than to the state.\vspace{-0.2cm}

\begin{definition}
\label{Definition_sync-state}Given the dynamical system in (\ref{linear
delayed system}), a solution $\{\mathbf{x}%
_{i}^{\star }(t)\}$ of (\ref{linear delayed system}) is said to be a
\emph{synchronized state} of the system, if
\begin{equation}
\dot{\mathbf{x}}_{i}^{\star }(t)=\mathbf{\omega }^{\star },\quad
\forall i=1,2,\ldots ,N.  \label{Def_synch_state}
\end{equation}%
The system (\ref{linear delayed system}) is said to \emph{globally}
synchronize if there exists a synchronized state as in
(\ref{Def_synch_state}), and \emph{all} the state derivatives
asymptotically converge to this common value, for \emph{any} given
set of initial conditions $\{{\mathbf{\phi }}_{i}\},$ i.e.,\vspace{-0.1cm}
\begin{equation}
\lim_{t\mapsto \infty }\Vert \dot{\mathbf{x}}_{i}(t)-\mathbf{\omega }^{\star
}\Vert =0,\,\,\,\,\,\,\,\forall i=1,2,\ldots ,N,  \label{def_eq}\vspace{-0.1cm}
\end{equation}%
where $\Vert \cdot \Vert $ denotes some vector norm and $\{\mathbf{x}%
_{i}(t)\}$ is a solution of (\ref{linear delayed system}). The
synchronized state is said to be \emph{globally asymptotically
stable} if the system globally synchronizes, in the sense specified in (\ref%
{def_eq}). The system (\ref{linear delayed system}) is said to
\emph{locally} synchronize if there exist disjoint subsets of the
nodes, called clusters, where the nodes in each cluster have state
derivatives converging, asymptotically, to the
same value, for any given set of initial conditions $\{{\mathbf{\phi }}%
_{i}\}$. \vspace{-0.1cm}
\end{definition}

According to Definition \ref{Definition_sync-state}, if there exists
a globally asymptotically stable synchronized state, then it must
necessarily be \emph{unique }(in the derivative). In the case of
local synchronization instead, the system may have multiple
synchronized clusters, each of them with a different synchronized
state. In the ensuing sections, we will provide necessary and
sufficient conditions for the system in (\ref{linear delayed
system}) to locally/globally synchronize, along with the closed form
expression of the synchronized state.\vspace{-0.3cm}

\section{Necessary and Sufficient Conditions for Self-Synchronization
\label{Sec:Sync-with-delays}}\vspace{-0.2cm}
To derive our main results, we rely on some basic notions of
digraphs theory, as briefly recalled next. A digraph $\mathscr{G}$
is defined as ${\mathscr{G}=}\{{\mathscr{V}, \mathscr{E}}\}$, where
$\mathscr{V}$ is the set of vertices and $\mathscr{E}\subseteq
\mathscr{V}\times \mathscr{V}$ is the set of edges, with the
convention that $e_{ij}=(v_i,v_j)\in \mathscr{E}$ if there exists an
edge from $v_j$ to $v_i$, i.e., the information flows from $v_j$ to
$v_i$. A digraph is weighted if a positive weight, denoted by
$a_{ij}$, is associated with each edge $e_{ij}$. The out-degree of a
vertex is defined as the sum of the weights of all its incoming
edges. The in-degree is similarly defined. The Laplacian matrix
$\mathbf{L}$ of a digraph is defined as
$\mathbf{L}=\mathbf{D}-\mathbf{A}$, where $\mathbf{D}$ is the
diagonal matrix of vertex out-degrees and $\mathbf{A}$ is the
adjacency matrix, defined as $[{A}]_{ij}=a_{ij}$. A digraph is a
directed tree if it has $N$ vertices and $N-1$ edges and there
exists a root vertex (i.e., a zero out-degree vertex) with directed
paths to all other vertices. A directed tree is a \emph{spanning}
directed tree of a digraph $\mathscr{G}$ if it has the same vertices
of $\mathscr{G}$. A forest is a collection of trees. A digraph is balanced if the out-degree of
each vertices is equal to its in-degree. A digraph is
\emph{strongly} connected (SC) if any ordered pair of distinct nodes
can be joined by a directed path. A digraph is \emph{quasi-strongly}
connected (QSC) if for every ordered pair of nodes $v_i$ and $v_j$
there exists a node $r$ that can reach both $v_i$ and $v_j$ by a
directed path. A digraph is \emph{weakly} connected (WC) if any
ordered pair of distinct nodes can be joined by a path, ignoring the
orientation of the edges.

The next theorem is the fundamental result of this paper and it
provides necessary and sufficient conditions for the proposed
decentralized approach to achieve local/global consensus in the
presence of propagation delays and nonreciprocal communication
links. \vspace{-0.2cm}

\begin{theorem}[\cite{Scutari-Barbarossa-Delay-J}]
\label{Theorem_delay-linear_stability}Let ${\mathscr{G}=}\{{%
%TCIMACRO{\TeXButton{V}{\mathscr{V}}}%
%BeginExpansion
\mathscr{V}%
%EndExpansion
,%
%TCIMACRO{\TeXButton{E}{\mathscr{E}}}%
%BeginExpansion
\mathscr{E}}\}$ be the digraph associated to the network in (\ref{linear delayed system}%
), with Laplacian matrix $\mathbf{L}$. Let $\mathbf{\gamma }=[\gamma
_{1},\ldots ,\gamma _{N}]^{T}$ be a left (normalized) eigenvector of $%
\mathbf{L}$ corresponding to the zero eigenvalue$,$ i.e., $\mathbf{\gamma }%
^{T}\mathbf{L}=\mathbf{0}_{N}^{T}$ and $\left\Vert \mathbf{\gamma }%
\right\Vert =1.$

Given the system in (\ref{linear delayed system}), assume that the
following conditions are satisfied: \textbf{a1}) The coupling gain
$K$ and the coefficients $\left\{ c_{i}\right\} $ are positive;
\textbf{a2}) The propagation delays $\{\tau _{ij}\}$ are finite,
i.e., $\tau _{ij}\leq \tau =\max_{i\neq j}\tau _{ij}<+\infty ,$
$\forall i\neq j;$ \textbf{a3}) The initial conditions are taken in
the set of continuous bounded
functions mapping the interval $[-\tau ,\ 0]$ to $%
%TCIMACRO{\U{211d} }%
%BeginExpansion
\mathbb{R}
%EndExpansion
^{N}.$

Then, system (\ref{linear delayed system}) globally synchronizes for
\emph{any} given set of propagation delays, if and only if the
digraph ${\mathscr{G}}$ is QSC. The synchronized state is given
by\vspace{-0.2cm}
\begin{equation}
\dot{x}_{q}^{\star }\triangleq \omega ^{\star
}=\frac{\dsum_{i=1}^{N}\gamma
_{i}c_{i}g_i(y_i)}{\dsum_{i=1}^{N}\gamma
_{i}c_{i}+K\dsum_{i=1}^{N}\dsum_{j\in \mathcal{N}_{i}}\gamma
_{i}a_{ij}\tau _{ij}},\,\,\forall q, \label{bias_Theo}
\end{equation}%
where $\gamma _{i}>0$ if and only if node $i$ can reach all the
other nodes of the graph through a directed path, otherwise $\gamma
_{i}=0$. \vspace{-0.2cm}
\end{theorem}

Theorem \ref{Theorem_delay-linear_stability} has a very broad
applicability, as it does not make any particular reference to the
network topology. If, conversely, the topology has a specific
structure, then we may have the following forms of
consensus.\footnote{We focus, w.l.o.g., only on WC digraphs. In the case of non WC digraphs, Corollary \ref{Corollary} applies to each disjoint component of the digraph.}\vspace{-0.3cm}

\begin{corollary}[\cite{Scutari-Barbarossa-Delay-J}]
\label{Corollary}Given system (\ref{linear delayed system}), assume
that conditions \textbf{a1-a3} of Theorem
\ref{Theorem_delay-linear_stability} are satisfied.
Then,\vspace{-0.1cm}

\begin{enumerate}
\item The system globally synchronizes to the state derivative%
\begin{equation}
\dot{x}_{q}^{\star }=g_r(y_r), \label{spanning-tree_sync_state}
\end{equation}%
$\forall q, r=1,\cdots ,N,$ if and only if the digraph
${\mathscr{G}}$ contains one spanning directed tree, with root node
given by node $r$.\vspace{-0.1cm}

\item The system globally synchronizes and the synchronized state is given
by (\ref{bias_Theo}) with all $\gamma _{i}$'s positive if and only
the digraph ${\mathscr{G}}$ is SC. The synchronized state becomes
\begin{equation}
\dot{x}_{q}^{\star }=\frac{\dsum_{i=1}^{N}c_{i}g_i(y_i)}{%
\dsum_{i=1}^{N}c_{i}+K\dsum_{i=1}^{N}\dsum_{j\in
\mathcal{N}_{i}}a_{ij}\tau _{ij}}, \label{bias_Theo_balanced_G}
\end{equation}%
$\forall q=1,\cdots ,N$, if and only if, in addition, the digraph
${\mathscr{G}}$ is balanced.

\item The system \emph{locally} synchronizes in $K$ disjoint clusters ${%
%TCIMACRO{\TeXButton{C}{\mathscr{C}}}%
%BeginExpansion
\mathscr{C}%
%EndExpansion
}_{1},\ldots ,$ ${%
%TCIMACRO{\TeXButton{C}{\mathscr{C}}}%
%BeginExpansion
\mathscr{C}%
%EndExpansion
}_{K}\subseteq \{1,\ldots ,N\}$,\footnote{In general, the clusters
$\mathscr{C}_{1},\ldots ,{\mathscr{C}}_{K}$ are not a partition of
the set of nodes $\{1,\cdots, N\}$.} with synchronized state
derivatives
\begin{equation}\label{Eq-Corollary-cluster}
\dot{x}_{q}^{\star }=\frac{\dsum_{i\in {%
%TCIMACRO{\TeXButton{C}{\mathscr{C}}}%
%BeginExpansion
\mathscr{C}%
%EndExpansion
}_{k}}\gamma _{i}c_{i}g_i(y_i)}{\dsum_{i\in {%
%TCIMACRO{\TeXButton{C}{\mathscr{C}}}%
%BeginExpansion
\mathscr{C}%
%EndExpansion
}_{k}}\gamma _{i}c_{i}+K\dsum_{i\in {%
%TCIMACRO{\TeXButton{C}{\mathscr{C}}}%
%BeginExpansion
\mathscr{C}%
%EndExpansion
}_{k}}\dsum_{j\in \mathcal{N}_{i}}\gamma _{i}a_{ij}\tau _{ij}},
%\quad \forall q\in {%
%%TCIMACRO{\TeXButton{C}{\mathscr{C}}}%
%%BeginExpansion
%\mathscr{C}%
%%EndExpansion
%}_{k},\, \forall k,
\end{equation}%
$\forall q\in {%
%TCIMACRO{\TeXButton{C}{\mathscr{C}}}%
%BeginExpansion
\mathscr{C}%
%EndExpansion
}_{k}$ and  $k=1,\cdots ,K,$ if and only if the digraph
${\mathscr{G}}$ is WC and it contains a forest with $K$ strongly
connected root components.
\end{enumerate}
\end{corollary}

\noindent \textbf{Remark 1: Robustness with respect to large
propagation delays. } The first important property of the proposed
system resulting from Theorem \ref{Theorem_delay-linear_stability}
is its robustness against propagation delays. It turns out in fact
that the convergence capability of system (\ref{linear delayed
system}) is not affected by the propagation delays. This property
represents the major
difference between our system and the scheme proposed in \cite%
{Olfati-Saber, Strogatz}, where, instead, even in the special case of \emph{%
homogeneous} delays (i.e., $\tau _{ij}=\tau ,$ $\forall i\neq j$) and \emph{%
undirected} connected graph (i.e., $a_{ij}=a_{ji},$ $\forall i\neq
j$), the average consensus is reached if and only if the common
delay $\tau $ is smaller than a topology-dependent threshold, whose
value is a decreasing function of the maximum graph degree.
 This implies, for example, that networks with hubs
(i.e., nodes with very large degrees) that are commonly encountered
in scale-free networks, are fragile against propagation delays,
under the protocol of \cite{Olfati-Saber, Strogatz}, even in the
simple case of homogeneous delays.

The reason for this difference in the convergence capabilities of
the two systems in the presence of propagation delays, is a
consequence of the alternative definition of global consensus that
we proposed for (\ref{linear delayed system}), with respect to the
classical one used in
\cite{Tsitsiklis-Bertsekas-Athans}$-$\cite{Olfati-Saber-Murray-ProcIEEE07} and
\cite{Strogatz}$-$\cite{Lee-Spong-06}. In fact, %differently from
%\cite{Olfati-Saber-2003}$-$\cite{Olfati-Saber-Murray-ProcIEEE07} and
%\cite{Strogatz}$-$\cite{Lee-Spong-06},
we do not require all the
state variables to converge to a common time-independent value, but
only to converge towards trajectories given by parallel straight
lines. %In other words, our system is not required to have a unique
%and global asymptotically stable equilibrium as in
%\cite{Olfati-Saber-2003}$-$\cite{Lucarelli04}.
 This extra
flexibility provides the additional degrees of freedom that make
possible the achievement of a consensus on the state derivative
without requiring any constraint on the propagation delays (besides
the obvious requirement of being bounded).\smallskip

\noindent \textbf{Remark 2: Effect of network topology on consensus structure. }%
Theorem \ref{Theorem_delay-linear_stability} generalizes all the
previous (only sufficient) conditions known in the literature
\cite[Ch. 7.3]{Tsitsiklis-Bertsekas-book}, \cite{Olfati-Saber}, \cite{Strogatz}$-$
\cite{Scutari-Barbarossa-Delay-SPAWC} for the convergence of linear
agreement protocols in the presence of propagation delays. In fact,
our theorem provides a full characterization of system (\ref{linear
delayed system}) in terms of necessary and sufficient conditions for
either global or local synchronization, valid for \emph{any}
possible degree of connectivity in the network (not only for SC
digraphs as in \cite{Olfati-Saber}, \cite{Strogatz}$-$
\cite{Lee-Spong-06}), as detailed next.\\
\indent In general, the digraph ${\mathscr{G}}$ modeling the
interaction among the nodes may have one of the following
structures: i) ${\mathscr{G}}$ contains only one directed spanning
tree, with a single root node, i.e., there exists only one node that
can reach all the other nodes in the network by a directed path; ii)
${\mathscr{G}}$ contains more than one directed spanning tree, i.e.,
there exist multiple nodes (possibly all the nodes), strongly
connected to each other, that can reach all the other nodes by a
directed path; iii) ${\mathscr{G}}$ is WC and contains a forest,
i.e., there exists no node that can reach all the others through a
directed path.
In the first two cases, according to Theorem \ref%
{Theorem_delay-linear_stability},  system (\ref%
{linear delayed system}) achieves a \emph{global} consensus, whereas
in the third case the system forms clusters of consensus with, in
general, different consensus values in each cluster, i.e., it
synchronizes only {\it locally}. In other words, global
synchronization is possible {\it if and only if} there exists at
least one node (the root node of the spanning directed tree of the
digraph) that can send its information, directly or indirectly, to
all other nodes. In particular,  if only one node can reach all the others, then the final consensus
will depend only on the observation taken from that node (see (\ref{spanning-tree_sync_state})). On the
other extreme, the global consensus contains contributions from
\emph{all} the nodes if and only if the graph is SC. If instead no node can reach all the others, then the information gathered
in each sensor has no way to propagate through the \emph{whole}
network and thus a global consensus cannot be reached. Still, a
\emph{local} consensus is
achievable among all the nodes that do influence each other (see (\ref{Eq-Corollary-cluster})).\\
%\indent Interestingly, the closed form expressions of the
%synchronized state as in (\ref{spanning-tree_sync_state}) confirms
%the above statements: The observation $y_i$ of, let us say, sensor
%$i$ affects the final consensus value if and only if such an
%information can reach all the other nodes by a strong directed path.
%As by-product
% of this result, we have the following special cases (Corollary 1).
%  If only one node can reach all the others, then the final consensus
%will depend only on the observation taken from that node. On the
%other extreme, the global consensus contains contributions from
%\emph{all} the nodes if and only if the graph is SC. \\
\indent As
an additional remark, the possibility to form clusters of consensus,
rather than a global consensus, depends on the channel coefficients
$a_{ij}$. These may be altered by changing the transmit powers
$P_j$. According to the previous comments, the nodes with the
highest transmit power will be the most influential ones. If, for
example, we want to write a certain value on each node, we can use
the same consensus mechanism used in this paper by assigning, for
example, that value to node $i$, and use transmit powers such that
node $i$ is the only node that can reach every other node.\smallskip

\noindent \textbf{Remark 3: Closed form expression of the
synchronized state. }An additional important  contribution of Theorem \ref%
{Theorem_delay-linear_stability} is to provide a closed form
expression of the synchronized state, as given in (\ref{bias_Theo}),
valid for any network topology (not only for undirected graphs as in
\cite{Olfati-Saber}). Expression (\ref{bias_Theo}) shows a
dependence of the synchronized state on the network topology and
propagation parameters, through the coefficients $\{a_{ij}\}$ and
the
delays $\{\tau _{ij}\}$.\\
Because of this dependence, the final consensus resulting from (\ref%
{bias_Theo}) cannot be made to coincide with the desired decision
statistics as given by (\ref{f}), except that in the ideal case
where all the delays are equal to zero. However, expression
(\ref{bias_Theo}) suggests also a method to get rid of the bias, as
shown in the following algorithm. We let the system in (\ref{linear
delayed system}) to evolve twice: the first time, the system evolves
according to (\ref{linear delayed system}) and we denote by
$\omega^*(\by)$ the synchronized state; the second time, we set
$g_i(y_i)=1$ in (\ref{linear delayed system}) and we let the system
evolve again, calling the final synchronized state $\omega(\buno)$.
From (\ref{bias_Theo}), if we take the ratio
$\omega^*(\by)/\omega^*(\buno)$,  we get\vspace{-0.2cm}
\begin{equation}
\frac{\omega^*(\by)}{\omega^*(\buno)}=%
\frac{\dsum_{i=1}^{N}\gamma _{i}c_{i}
g_{i}(y_i)}{\dsum_{i=1}^{N}\gamma _{i}c_{i}},  \label{ratio}
\end{equation}%
which coincides with the ideal value achievable in the absence of
delays.\\Thus, using this simple two-step algorithm, we are able to
control the value of the synchronized state in advance, without
affecting the convergence capabilities of the system. This makes our
system strongly different from the linear agreement protocols
proposed in \cite[Ch. 7.3]{Tsitsiklis-Bertsekas-book}, \cite{Papachristodoulou-CDC06, Lee-Spong-06}. In fact,
the dynamical systems studied in \cite[Ch. 7.3]{Tsitsiklis-Bertsekas-book}, \cite{Papachristodoulou-CDC06,
Lee-Spong-06} are guaranteed to reach an agreement for any set of
nonhomogeneous delays (provided that the digraph is QSC in \cite[Ch. 7.3]{Tsitsiklis-Bertsekas-book} and SC in \cite{Papachristodoulou-CDC06,
Lee-Spong-06}), but this
common value is not related to the sensors' measurements by a known
function. In other words, the global
consensus asymptotically achieved by the protocols in \cite[Ch. 7.3]{Tsitsiklis-Bertsekas-book} and \cite%
{Papachristodoulou-CDC06, Lee-Spong-06} is a priori unpredictable,
so that the systems proposed in the cited works
cannot be used directly to distributively compute prescribed decision tests
or sufficient statistics of sensors' measurements, as given in
(\ref{f}). \vspace{-0.2cm}
\section{Numerical Results and Conclusions}\label{Sec:Numerical-Results} \vspace{-0.2cm}In this section, we illustrate
first some examples of consensus, for different network topologies.
Then, we show an application of the proposed technique to an
estimation problem, in the presence of random link
coefficients.\smallskip

\begin{figure}[tbh]
\hspace{-0.7cm}\includegraphics[height=7.3 cm]{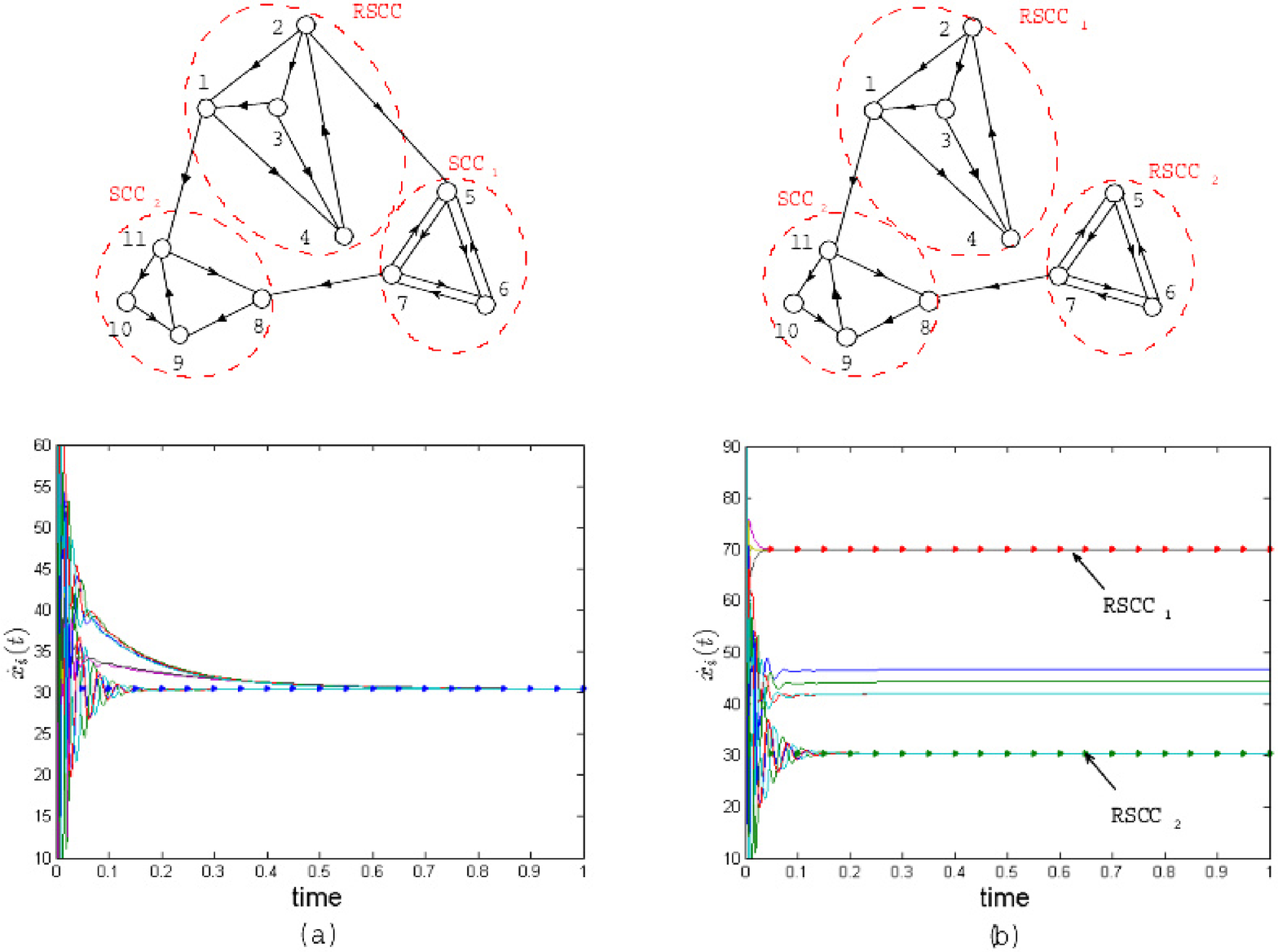}\vspace{-0.5cm}
\caption{\footnotesize Self-synchronization for two different
network topologies: a) QSC digraph with three SCCs; b) WC digraph
with a two trees forest; $T_s=10^{-3}$ s, $\tau=50 T_s$, $K=30$.
}\vspace{-0.5cm} \label{Figure-RSCC}
\end{figure}

 \noindent \textbf{Example 1:
Different forms of consensus for different topologies}\\ In Figure
\ref{Figure-RSCC}, we consider two topologies  (top row), namely:
(a) a QSC digraph, (b) a WC (not QSC) digraph with a forest composed
by two trees. For each digraph, we also sketch its decomposition
into Strongly Connected Components (SCCs) (each one enclosed in a
circle), and we denote by RSCC the root SCC of any spanning directed
tree contained in the digraph. In the bottom row of Figure
\ref{Figure-RSCC}, we plot the dynamical evolutions of the state
derivatives of system (\ref{linear delayed system}) versus time, for
the two network topologies,
together with the theoretical asymptotic values predicted by (\ref%
{bias_Theo}) (dashed line with arrows). As proved by Theorem \ref%
{Theorem_delay-linear_stability}, the dynamical system in Figure \ref%
{Figure-RSCC}a) achieves a global consensus, %since %the underlying
%digraph is SC. The network of Figure \ref{Figure-RSCC}b), instead,
%is not SC, but the system is still able to globally synchronize,
since there is a set of nodes (those in the RSCC component) able to
reach all other nodes. The final consensus contains only the
contributions of the nodes in the RSCC, since no other node belongs
to the root SCC of a spanning directed
tree of the digraph. The system in Figure \ref%
{Figure-RSCC}b) cannot reach a global consensus since there is no
node that can reach all the others, but it does admit two disjoint
clusters, corresponding to the two RSCCs, namely RSCC$_1$ and
RSCC$_2$. The middle lines of Figure \ref {Figure-RSCC}b) refer to
the nodes of the SCC component, not belonging to either RSCC$_1$ or
RSCC$_2$, that are affected by the consensus achieved in the two
RSCC components, but that cannot affect them. Observe that, in all
the cases, the state derivatives of the (global or local) clusters
converge to the values predicted by the closed form expression given
in (\ref{bias_Theo})-(\ref{Eq-Corollary-cluster}), depending on the
network topology.\smallskip

\noindent \textbf{Example 2}: \textbf{Distributed optimal decisions
through} \textbf{consensus}. The behaviors shown in the previous
example refer to a given realization of the topology, with given
link coefficients, and of the observations. In this example, we
report a global parameter representing the variance obtained in the
estimate of a scalar variable. Each sensor observes a variable $y_i=
A_i \xi+w_i$, where $w_i$ is additive zero mean Gaussian noise, with
variance $\sigma_i^2$. The goal is to estimate $\xi$. The estimate
is performed through the interaction system (\ref{linear delayed
system}), with functions $g_i(y_i)=y_i/A_i$ and coefficients
$c_i=A_i^2/\sigma_i^2$, chosen in order to achieve the globally
optimal ML estimate. The network is composed of $40$ nodes, randomly
spaced over a square of size $D$. We set the threshold on the
amplitude of the minimum useful signal so that the underlying
digraph be SC. The analog system (\ref{linear delayed system}) is
implemented in discrete time, with sampling time $T_s=10^{-3}$ sec.
The size of the square occupied by the network is chosen in order to
have a maximum delay $\tau=100 T_s$. To simulate a practical
scenario, the channel coefficients $a_{ij}$ are generated as i.i.d.
Rayleigh random variables, to accommodate for channel fading. Each
variable $a_{ij}$ has a variance depending on the distance $d_{ij}$
between nodes $i$ and $j$, equal to\footnote{We use the attenuation
factor $1/(1+d_{ij}^2)$ instead of $1/d_{ij}^2$ to avoid the
undesired event that, for $d_{ij}<1$ the received power might be
greater than the transmitted power.}
$\sigma_{ij}^2=P_j/(1+d_{ij}^2)$.

\begin{figure}[tbh]
\vspace{-0.5cm}\centering\includegraphics[height=5.4cm]{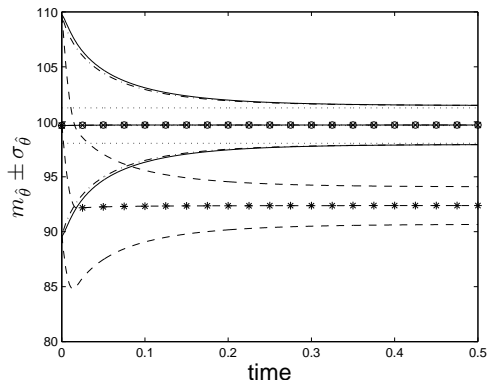}\vspace{-0.5cm}
\caption{{\protect\footnotesize Estimated parameter vs. convergence
time.}} \label{Figure-variance-vs-time}\vspace{-0.3cm}
\end{figure}

In Figure \ref{Figure-variance-vs-time}, we plot the estimated
average state derivative (plus and minus the estimation standard
deviation), as a function of the iteration time. The averages are
taken over the nodes, for $100$ independent realizations of the
network, where, in each realization we generated a new topology and
a new set of channel coefficients and noise terms. The results refer
to following cases of interest: a) ML estimate achieved with a
centralized system, with no communication errors between nodes and
fusion center (dotted lines); b) estimate achieved with the proposed
method, with no propagation delays, as a benchmark term (dashed and
dotted lines plus $\times$ marks for the average value); c) estimate
achieved with the proposed method, with propagation delays (dashed
lines plus stars for the average value); d) estimate achieved with
the two-step estimation method leading to (\ref{ratio}) (solid lines
plus circles for the average value).
From Figure \ref{Figure-variance-vs-time}, we can see that, in the
absence of delays, the (decentralized) iterative algorithm based on
(\ref{linear delayed system}) behaves, asymptotically, as the
(centralized) globally optimal ML estimator. In the presence of
delays, we observe a clear bias (dashed lines), due to the large
delay values, but still with a final estimation variance close to
the ML estimator's. Interestingly, if the two-step procedure leading
to (\ref{ratio}) provides results very close to the optimal ML
estimator, with no apparent bias, in spite of the large delays and
the random channel fading coefficients.

In conclusion, in this paper we have proposed a totally
decentralized sensor network scheme capable to reach globally
optimal decision tests through local exchange of information among
the nodes, in the presence of nonreciprocal communication channels
and inhomogeneous time-invariant propagation delays. %The method is
%particularly useful for applications where the goal of the network
%is to take decisions about one common phenomenon.
 Differently from
the average consensus protocols available in the literature, our
system globally synchronizes for {\it any} set of (finite)
propagation delays \emph{if and only if} the underlying digraph is
QSC, with a final synchronized state that is a \emph{known} function
of the sensor measurements. In general, the synchronized state
depends on the propagation parameters, such as delays and the
communication channels. Nevertheless, exploiting our closed form
expression for the final consensus values, we have shown how to
recover an unbiased estimate, for any realization of delays and
channel coefficients, without the need to know or estimate these
coefficients. If we couple the nice properties mentioned above with
the properties reported in \cite{Barbarossa-Scutari-Journal}, where
we showed that, in the absence of delays, the consensus protocol
proposed in this paper and in \cite{Barbarossa-Scutari-Journal} is
also robust against coupling noise, we have, overall, a good
candidate for a distributed sensor network.\vspace{-0.2cm}

\input{biblio_revised.tex}

\end{document}

%% file: biblio_revised.tex
%\vspace{-0.2cm}
\def\baselinestretch{.1}